\documentstyle[12pt]{article}
\newfont{\bg}{cmr10 scaled\magstep4}

\newcommand{\bigzerou}{%
   \smash{\lower1.7ex\hbox{\bg 0}}}
\begin{document}
\title{Mirror Symmetry on Arbitrary Dimensional
Calabi-Yau Manifold with a few moduli}
\author{ Masaru Nagura\thanks{A fellow of t[he Japan Society
for the Promotion of Science for Japanese Junior Scientists }
\thanks{nagura@danjuro.phys.s.u-tokyo.ac.jp}
\\
{\it Department of Phisics, The University of Tokyo}\\
{\it  Bunkyo-ku, Tokyo 113, Japan}}
\maketitle

\begin{abstract}
We calculate the B-model on the mirror pair of
$X_{2N-2}(2,2,\cdots,2,1,1)$ ,
which is an $(N-2)$-dimensional Calabi-Yau manifold and
has two marginal operators i.e. $h^{1,1}(X_{2N-2}(2,2,\cdots,2,1,1))=2$.
In \cite{nagandjin} we have discussed about mirror symmetry on
$X_N(1,1,\cdots,1)$ and its mirror pair.
However, $X_N(1,1,\cdots,1)$ had  only one moduli.
In this paper we extend its methods
to the case with a few moduli
using toric geometry.
\end{abstract}
\section{Introduction}
In \cite{nagandjin}
Jinzenji and myself discussed about mirror symmetry
of arbtrary dimensional case.
In \cite{hosono}
Hosono et.al., on the other hand, established a method dealing with
mirror symmetry with a few moduli.
But they delt with the only case of 3-dimension.
We want to unite these methods , i.e. to consider arbitrary dimensional
Calabi-Yau manifold with a few moduli.
We will realize that idea especially in the case of
$X_{2N-2}(2,2,\cdots,2,1,1)$. But it will be appearent later that
our method is valid in other models such as
$X_{4N-8}(2N-4,2,2,\cdots,2,1,1),\;
X_{4N-6}(2N-3,2,2,\cdots,2,1),\;
X_{6(N-3)}(3(N-3),N-3,N-3,1,\cdots,1),\;
X_{4(N-3)}(N-3,N-3,N-3,1,\cdots,1)$
and so on \cite{ZOKUHEN}.
\section{Toric data for $X_{2N-2}(2,2,\cdots,2,1,1)$ }
The method of toric geometry is the most suitable way
to deal with Calabi-Yau manifolds and their embedding spaces systematically.

The integral convex polyhedron which is associated with
$X_{2N-2}(2,2,\cdots,2,1,1)$ is the convex hull of the integral vectors
in $R^{N-1}$:
\begin{eqnarray}
& & \nu_1 =(N-2,-1,\cdots,-1,-1),\;\;
\nu_2 =(-1,N-2,-1,\cdots,-1,-1),\\ \nonumber
& & \;\;\cdots
\nu_{N-1} =(-1,-1,-1,\cdots,-1,N-2),\\  \nonumber
& & \nu_N =(-1,-1,-1,\cdots,-1).
\end{eqnarray}
For later convenience we call it $\Delta$.
A complete rational fan $\Sigma (\Delta)$ is associated to $\Delta$.
The toric variety $P_{\Delta}$ is defined as the toric variety associated
to the fan $\Sigma (\Delta)$ .In our case, of course, $P_{\Delta}=
P_{\Sigma (\Delta)}=P^{N-1}(2,2,\cdots,2,1,1)$. (For detail see \cite{oda}.)
Furthermore we naturally associate to $\Delta$ a Calabi-Yau hypersurface
which is defined as the zero locus $Z_f$ of the following Laurent polynomial
\begin{equation}
f(a,X)=\sum_i a_i X^{\nu_i},
\end{equation}
here $X^{\nu}=X_1^{\nu_1}\cdots X_{N-1}^{\nu_{N-1}}$
and $\nu_i$ runs over all the integral points in and on $\Delta$.
The integral convex polyhedron which is associated with its mirror pair,
let us call it $\Delta^*$, is also a convex hull of  integral vectors:
\begin{eqnarray}
 & & \nu_1^* =(1,0,\cdots,0,0),\;\;
\nu_2^* =(0,1,\cdots,0,0),\;\;\cdots
\nu_{N-1}^* =(0,0,\cdots,0,1),\\ \nonumber
 & & \nu_N^* =(-2,-2,\cdots,-2,-1).
\end{eqnarray}
Fortunately there are
always only two integral points which lie on the face of $\Delta^*$
or its interior.
\begin{eqnarray}
\nu_0^* =(0,0,\cdots,0,0),\\ \nonumber
\nu_{N+1}^* =(-1,-1,\cdots,-1,0).
\end{eqnarray}
note that $\nu_{N+1}^*=(-1,-1,\cdots,-1,0)$ is on the edge of
codimension $(N-2)$. Thus
\begin{equation}
h^{1,1}(\Delta)=h^{N-3,1}(\Delta^*)=2.
\end{equation}
The A-model associated with  $\Delta$ the B-model associated with $\Delta^*$
is 2-parameter system of $(N-2)$-dimension.
This is the simplest model suitable for our aim.

\section{The Picard-Fuchs differential equation}
Let us introduce $\bar{\nu}_i = (1,\nu_i) \in R^N $
for each $ \nu_i \in R^{N-1}$.
And let us define $L$ as
\begin{equation}
L=\{ (l_0,l_1,\cdots,l_N)\in Z^{N+1}| \sum_{i=0}^{N}l_i\bar{v}_i=0\}.
\label{eqL}
\end{equation}
By an integral basis $\{l_i\}$ $L$ can be expanded as follows
\begin{equation}
L=Zl^{(1)}\oplus\cdots\oplus Z{l^{(h)}}.
\end{equation}
where $h=h^{1,1}(\Delta)=h^{N-3,1}(\Delta^*)$.
According to Batyrev \cite{batyrev} the Picard-Fuchs defferential
equations for the Calabi-Yau space which is naturally defined in
$P_{\Delta^*}$ as zero locus of the polynomial $f=\sum a_i X^{\nu_i^*}=0$
are constructed in terms of $L=Zl^{(1)}\oplus\cdots\oplus Z{l^{(h)}}$.
For any $l^{(k)}\in L$, we obtain the Picard-Fuchs equations;
\begin{equation}
\left\{
\prod_{l_i^{(k)}}
\left(
\prod_{i=0}^{l_i^{(k)}-1}(\Theta_j-i)
\right)
-
\prod_{i=1}^{|l_0^{(k)}|}(i-|l_0^{(k)}|-\Theta_0)
\prod_{l_i^{(k)},j\neq 0}
\left(
\prod_{i=0}^{|l_i^{(k)}|-1}(\Theta_j+|l_i^{(k)}|-i)
\right)
x_k
\right\}
\Pi(x)=0
\end{equation}
where
\begin{equation}
x_k=(-1)^{l_0^{(k)}}a^{l^{(k)}},
\label{eq5}
\end{equation}
and $$\Theta_j=\sum_{k=1}^{h}l_j^{(k)}x_k\frac{\partial}{\partial x_k}.$$
However the integral basis for $L$ is not always unique.
We choose the basis $\{ l_i\}$ to be the basis of Mori cone
which is unique in the case of $X_{2N-2}(2,2,\cdots,2,1,1)$
\footnote{The number of Mori cone corresponds to the numberof
desingularization of $P_{\Delta^*}\;(P_{\Delta})$}, since such
a basis is associated to a good coordinate via (\ref{eq5})
to describe the large radius limit
structure.
For the case of $X_{2N-2}(2,2,\cdots,2,1,1)$ we have a basis of Mori cone;
\begin{eqnarray}
l^{(1)} = (-(N-1),1,\cdots,1,0,0,1),\\ \nonumber
l^{(2)} = (0,0,\cdots,0,1,1,-2).
\end{eqnarray}
Then we have two Picard-Fuchs defferential equations.
For $l^{(1)}$
\begin{equation}
\theta_x^{N-3}(\theta_x-\theta_y)-(N-1)x((N-1)\theta_x+(N-2))
((N-1)\theta_x+(N-3))\cdots ((N-1)\theta_x+1)\Pi=0,
\label{eq6}
\end{equation}
and for $l^{(2)}$
\begin{equation}
\theta_y^2-y(2\theta_y-\theta_x+1)(2\theta_y-\theta_x)\Pi=0,
\label{eq7}
\end{equation}
where $$ x=x_1,\;y=x_2,$$ and$$\theta_x=x\frac{\partial}{\partial x},\;
\theta_y=y\frac{\partial}{\partial y}.$$
\section{$(N-2)$-point Yukawa coupling}
Since there are two marginal operators, we have $(N-1)$ kinds of
$(N-2)$-point Yukawa coupling. Let $W^{i,j}$ be
\begin{equation}
W^{(i,N-2-i)}\equiv \int \Omega \wedge
(\bar{x}\frac{\partial}{\partial\bar{x}})^i
(\bar{y}\frac{\partial}{\partial\bar{y}})^{N-2-i}
\Omega,
\end{equation}
where $$\bar{x}=(N-1)^{(N-1)}x_1,\;\bar{y}=4x_2.$$
{}From the Picard-Fuchs equation (\ref{eq6}) (\ref{eq7}) and the fact that
$W^{(i,j)}=0$ for $i+j<N-2$ we obtain a recursion relation that
\begin{equation}
(1-\bar{y})W^{(N-4-i,i+2)}+\bar{y}W^{(N-3-i,i+1)}-\bar{y}/4W^{(N-2-i,i)}=0.
\end{equation}
Then $W^{(N-2-i,i)}$ can be expressed as
$$
W^{(N-2,0)} \times ( Rational\; expression\;\;of \;\;\bar{x},\;\;\bar{y}).
$$
{}From $\theta_x$(\ref{eq6}) and $\theta_y$(\ref{eq7})
and trivial relations:
\begin{eqnarray*}
W^{(N-1,0)}& = & \frac{N-1}{2}(x\partial_x)W^{(N-2,0)},\\
W^{(N-2,1)}& = & \frac{N-2}{2}(x\partial_x)W^{(N-3,1)}
+\frac{1}{2}(y\partial_y)W^{(N-2,0)},\\
W^{(N-3,2)}& = & \frac{N-3}{2}(x\partial_x)W^{(N-4,2)}
+\frac{2}{2}(y\partial_y)W^{(N-3,1)},
\end{eqnarray*}
it turns out
\begin{equation}
W^{(N-2,0)}=\{(1-\bar{x})^2-\bar{x}^2 \bar{y}\}^{-1} \times constant.
\label{eq15}
\end{equation}
Thus we can express all the $(N-2)$-point Yukawa couplings as
rational expressions of $\bar{x},\bar{y} $ up to constant.
\section{Transformation to the A-model}
Obviously the Picard-Fuchs equation (\ref{eq6}) and (\ref{eq7}) are maximally
unipotent at $x=y=0$.
Regular solution around $x=y=0$ is
\begin{equation}
\omega_0(x,y)=\sum_{n_1,n_2 \geq 0}
\frac
{\{n_1(N-1)\}!}
{(n_1!)^{N-2}(n_2!)^2(n_1-2n_2)!}
x^{n_1}y^{n_2}.
\end{equation}
Solutions which have singularity of $\log x$ or $\log y$ at $x=y=0$
are
\begin{eqnarray}
\lefteqn{\omega_x(x,y)=\omega_0(x,y)\log x}\\ \nonumber
& & +\sum_{n_1,n_2 \geq 0}\{(N-1)\Psi((N-1)n_1+1)-(N-2)\Psi(n_1+1)\\ \nonumber
& &- \Psi((n_1-2n_2+1)\}c(n) x^{n_1} y^{n_2}, \\
\lefteqn{\omega_y(x,y)=\omega_0(x,y)\log y}\\ \nonumber
& & +\sum_{n_1,n_2 \geq 0}\{-2\Psi(n_2+1)+2\Psi(n_1-2n_2+1)\}
c(n) x^{n_1} y^{n_2},
\end{eqnarray}
where
$$
c(n_1,n_2)= \frac
{\{n_1(N-1)\}!}
{(n_1!)^{N-2}(n_2!)^2(n_1-2n_2)!}
$$
The mirror maps are given by
\begin{eqnarray}
t_x=\omega_x/\omega_0\\
\label{eq16}
(t_y=\omega_y/\omega_0).\nonumber
\end{eqnarray}
But these mirror maps are not nessesarily equal to the variable $\tilde{t}_i$
associated with the integral cohomology basis $h_i\in H^{1,1}(X,Z)$ .
In \cite{hosono} Hosono et.al. solved this ploblem using several ansatzs.
At first let the relation between $t_i$ and $\tilde{t}_i$ be linear;
\begin{equation}
t_i=\sum_j m_{ij}\tilde{t}_j ,\;\;m_{ij}\in Z
\end{equation}
We need some knowledge of K{\"a}hler cone to determine $m_{ij}$.
K{\"a}hler cone $K_u$ is identified with the set of the strictly convex
piecewise lininear
functions $u$ which is defined in a convex hull of the origin and the set
$(1,\Delta^*)$.
Using the basis of Mori cone $\{l^{(i)}\}$ the inequalities which is
equivalent to the very condition that $u$ is strictly convex is expressed
as $<u,l^{(i)}> > 0$. Furthermore in terms of $\{l^{(i)}\}$, $K_u$
can be written as
\begin{equation}
K_u=\sum_i c_i
\left(
\sum_k n_{ik}< u,l^{(k)}>
\right)
e_{\nu_i^*},
\end{equation}
where $n_{ij}\in Z,c_i\in Q$.
(See Appendix for detail.)

On the other hand, from (\ref{eq16}) we have in the large radius limit
$ X_k \to 0$,
\begin{equation}
t_k \sim \log x_k \sim \sum_i (\log a_i) l_i^{(k)} \ll 0.
\end{equation}
{}From this observation we postulate the asymptotic relation that
$u_i=\log a_i$. Thus we have
$$ K_u=\sum_i \sum_{j,k}^h c_i n_{ij} m_{jk} \tilde{t}_k e_{\nu_i^*}.$$
Furthermore we postulate the r.h.s. should be
\begin{equation}
=\sum_i \tilde{c}_i \tilde{t}_i e_{\nu_i^*}
{}.
\end{equation}
Then
$$
m=
\left(
\begin{array}{cc}
(N-1)\tilde{c}_1/c_1 & -\tilde{c}_2/c_2 \\
0                    & 2\tilde{c}_2/c_2
\end{array}
\right).
$$
{}From the fact of algebraic geomertry that
\begin{eqnarray*}
J\cdot J \cdots J\cdot J & = & 2^{N-2},\\
J\cdot J \cdots J\cdot D & = & 0,\\
J\cdot J \cdots J\cdot D \cdot D & = & -2^{N-2},\\
\vdots
\end{eqnarray*}
where J is the canonical diviser on $X_{2N-2}(2,2,\cdots,2,1,1)$
and D is the exceptional one,
the (N-2)-point Yukawa coupling should have the following classical
asymptotic feature $(x_k \to 0)$
such that
\begin{eqnarray*}
W_{\tilde{t}_x \cdots \tilde{t}_x }
& = &  2^{N-2}+O(e^t),\\
W_{\tilde{t}_x \cdots \tilde{t}_x \tilde{t}_y}
& = & 0 +O(e^t),\\
W_{\tilde{t}_x \cdots \tilde{t}_x \tilde{t}_y \tilde{t}_y}
& = & -2^{N-2} +O(e^t),\\
& \vdots & \;\;.
\end{eqnarray*}
Then
$$
m=
\left(
\begin{array}{cc}
c & c \\
0 & -c
\end{array}
\right) ,\;\;
c\; is \; a\; constant.
$$
This $"c"$ is corresponding to the constant in (\ref{eq15}).
{}From the requirement that the coefficient of higher degree
of the Yukawa couplings should be integral, we finaliy obtain $c=1$ :
\begin{equation}
m=
\left(
\begin{array}{cc}
1 & 1 \\
0 & -2
\end{array}
\right).
\end{equation}
Now we can carry out the calculation of the (N-2)-point Yukawa coupling
of the A-model on $X_{2N-2}(2,2,\cdots 2,1,1)$;
\begin{eqnarray}
\lefteqn{W_{\tilde{t}_x \cdots \tilde{t}_x \tilde{t}_y \cdots \tilde{t}_y}
=}\\ \nonumber
& & \frac{1}{\omega_0(x(\tilde{t})^2}
 \sum_{i,\cdots, j,k,\cdots,l}
\frac{\partial x_i(\tilde{t})}{\partial \tilde{t}_x}
\cdots \frac{\partial x_j(\tilde{t})}{\partial \tilde{t}_x}
\frac{\partial x_k(\tilde{t})}{\partial \tilde{t}_k}
\cdots \frac{\partial x_l(\tilde{t})}{\partial \tilde{t}_y}
\frac{1}
{x_i(\tilde{t}) \cdots x_j(\tilde{t}) x_k(\tilde{t}) \cdots x_l(\tilde{t})}
W_{x_i \cdots x_j x_k \cdots x_l}
\end{eqnarray}
\section*{Conclusion}
We considered $X_{2N-2}(2,2,\cdots,2,1,1)$ and its mirror pair
in terms of toric geometry.
We obtained  the concrete formula of the (N-2)-point correlation functions
and the mirror map.
We should also remark that this method is valid in another model
\cite{ZOKUHEN}.
\section*{Appendix: Strictly convex piecewise linear function and
K\"{a}hler cone}
Oda and Park \cite{oda2} have shown how to calculate the k\"{a}hler cone
of $P_{\Delta}$ based on the toric data encoded in the polyhedron $\Delta$.

Consider the dual polyhedron $\Delta^*$ of $\Delta$.
(Suppose they are (N-2) dimensional for later convenience.)
And extend $\Delta^*$ to the polyhedron $\bar{\Delta}^*$ ($\in R^{N-1}$)
which is a convex hull of the origin and the set $(1, \Delta^*)$.
Let $\bar{\Xi}$ be the set of integral points in $\bar{\Delta}^*$
which is not inside a codimension one face of $\bar{\Delta}^*$.
Next let us imagine a simplicial decomposition of $\bar{\Delta}^*$
induced by a simplicial decomposition of $\Delta^*$.
A piecewise linear function $u$ is determined by assiging real values
$u_j$ to each integral point $\bar{\nu_j}^* \in \bar{\Xi}$.
If vertices of a (N-1)dimensional simplex $\sigma_i$ are given by
$\bar{\nu}_{i_1^*},\cdots ,\bar{\nu}_{i_{N-1}^*} \in \bar{\Xi}$,
then the piecewise linear function $u$ restricted in $\sigma$ is defined as
\begin{equation}
u(v)= c_{i_1} u_{i_1}+ \cdots + c_{i_{N-1}} u_{i_{N-1}},
\end{equation}
where $v$ is an arbitrary point in $\sigma_i$ which is written as
$v=c_{i_1} \bar{\nu_{i_1}}^* + \cdots + c_{i_{N-1}}  \bar{\nu_{i_{N-1}}}^*$.

Equivalently, the piecewise linear function $u$ can be described by a set of
vectors $Z_{\sigma}\in Z^{N-1}$ such that
\begin{equation}
u(v)=<Z_{\sigma}, v> ,\;\;v\in \sigma,
\end{equation}
where $\sigma$ runs in the set of the simplex of $\bar{\Delta}^*$.
A strictly convex piecewise linear function $u$ is
a piecewise linear function $u$ with the property;
\begin{eqnarray}
u(v)=<Z_{\sigma}, v> ,\;\;for\;v\in \sigma,\\ \nonumber
u(v) > <Z_{\sigma}, v> ,\;\;for\;v\in \sigma.
\label{eq28}
\end{eqnarray}
The set of the strictly convex piecewise linear functions has a structure
of a cone, since if $u$ is a strictly convex piecewise linear function
then so is $\lambda u$ for $\lambda > 0$.

According to Oda and park, the strictly convex piecewise linear functions
constitute a cone in a quatient space
\begin{eqnarray}
V'={W'}_1 / \{\sum_{\xi \in \Xi}
<x,\bar{\xi}> e_{\xi}|x\in R^{N-1}\},\\ \nonumber
{W'}_1=\sum_{\xi \in \Xi}Re_{\xi},
\end{eqnarray}
where $e_{\xi}$ is the basis of the vector space ${W'}_1$, $\Xi$ is
the set of integral points in $\Delta$ which is not inside
a codimension one face of $\Delta$ and $\bar{\xi}=(1,\xi)$.
If we restrict $V'$ to the region satisfying the inequalities (\ref{eq28}),
we have a cone.
The k\"{a}hler cone is identifyed with this cone.

In the case of $X_{2N-2}(2,2,\cdots,2,1,1)$,
the $\bar{\Delta}^*$ decompose into (2N-2) simplexes.
(2N-2) simplexs consists of
$\sigma_{1,N},\cdots,\sigma_{N-2,N},
\sigma_{1,N-1},\cdots,\sigma_{N-2,N-1},\sigma_{N-1,N+1}$ and $ \sigma_{N,N+1}$,
where $\sigma_{i,j}=<0,\bar{\nu}_0^*,\cdots,\check{\bar{\nu}}_i^*,
\cdots,\check{\bar{\nu}}_j^*,\cdots,\bar{\nu}_{N+1}^*>$.
Then applying (\ref{eq28}) to these simplexes, we obtain two independent
inequalities
\begin{eqnarray}
-(N-1)u_0 + u_1 + u_2 + \cdots + u_{N-2} + u_{N+1} & > & 0 , \\ \nonumber
u_{N-1}+u_N-2u_{N+1} & > & 0,
\end{eqnarray}
If we write these inequalities in the form $<u,l^{(k)}>>0$, the $l^{(k)}$
form a paticular basis for the lattice of $L$ of the points $\bar{\Xi}$
(see (\ref{eqL})). This basis generates the cone called {\it Mori cone}.
These inequalities produce the K\"{a}hler cone for
$X_{2N-2}(2,2,\cdots,2,1,1)$.
\begin{eqnarray}
\lefteqn{K_u \equiv - \sum u_i e_{\nu_i^*}}\\ \nonumber
& & mod(e_{\nu_0^*}+e_{\nu_1^*}+\cdots+e_{\nu_N^*}+e_{\nu_{N+1}^*}=0,\\
\nonumber
& & e_{\nu_1^*}-2e_{\nu_N^*}-e_{\nu_{N+1}^*}=0,
\cdots,
e_{\nu_{N-2}^*}-2e_{\nu_N^*}-e_{\nu_{N+1}^*}=0,
e_{\nu_{N-1}^*}-e_{\nu_N^*}=0
) \\ \nonumber
& & =\frac{1}{2(N-1)}\{-2(N-1)u_0 +2(u_1+u_2+\cdots+u_{N-2})
+u_{N-1}+u_N \}e_{\nu_0^*}+\\ \nonumber
& & \frac{1}{2}(u_{N-1}+u_N-2u_{N+1})e_{\nu_{N+1}^*}
\end{eqnarray}
If we write the above using Mori cone's basis $\{l^{(k)}\}\;k=1,2$,
we obtain
\begin{equation}
K_u=\frac{1}{2(N-1)}<u,2l^{(1)}+l^{(2)}>e_{/nu_0^*}+
\frac{1}{2}<u,l^{(2)}>e_{\nu_{N+1}^*}.
\end{equation}

\end{document}